\documentclass[twocolumn,showpacs,preprintnumbers,amsmath,amssymb]{revtex4}

\usepackage{graphicx}
\usepackage{dcolumn}
\usepackage{bm}
\usepackage[pagebackref]{hyperref}

\begin{document}

\preprint{APS/123-QED}

\title{Solid-fluid transition in a granular shear flow}

\author{Ashish V. Orpe}
\author{D. V. Khakhar}%
\email{khakhar@iitb.ac.in} \affiliation{Department of Chemical
  Engineering, Indian Institute of Technology - Bombay, Powai, Mumbai,
  400076, India}

\date{\today}

\begin{abstract}
  
  The rheology of a granular shear flow is studied in a quasi-2d
  rotating cylinder.  Measurements are carried out near the midpoint
  along the length of the surface flowing layer where the flow is
  steady and non-accelerating. Streakline photography and image
  analysis are used to obtain particle velocities and positions.
  Different particle sizes and rotational speeds are considered. We
  find a sharp transition in the apparent viscosity ($\eta$) variation
  with rms velocity ($u$). In the fluid-like region above the depth
  corresponding to the transition point (higher rms velocities) there
  is a rapid increase in viscosity with decreasing rms velocity. Below
  the transition depth we find $\eta \propto u^{-1.5}$ for all the different
  cases studied and the material approaches an amorphous solid-like
  state deep in the layer. The velocity distribution is Maxwellian
  above the transition point and a Poisson velocity distribution is
  obtained deep in the layer. The observed transition appears to be
  analogous to a glass transition.

\end{abstract}

\maketitle

Granular materials are known to exist in solid-like and fluid-like
states \citep{jag96}.  Physical understanding of the flow of granular
materials has thus developed along two major themes based on the flow
regime \citep{jack86}. In the {\it rapid flow \/}| fluid-like |
regime, both theory and experimental analysis are generally cast in
the framework of the kinetic theory \citep{cam90}. In contrast, the
{\it slow flow \/}| solid-like | regime is most commonly described
using the tools of soil mechanics and plasticity theory \citep{ned92}
and recently by analogy to glasses \citep{danna01a,bar01,kurch00}.  A
key difference between solid-like and fluid-like states, which appears
to be emerging from recent studies, is the definition of a temperature
in the two states. In the fluid-like state the granular temperature is
defined as the kinetic energy of velocity fluctuations in analogy with
the kinetic theory of gases. In contrast, for dense granular flows
(solid-like state), numerical \citep{mak02b} and experimental
\citep{danna03b} studies show the validity of the
fluctuation-dissipation theorem (the mean-square displacement of a
particle is proportional to time for a constant applied force). This
allows for the definition of a new temperature \citep{kurch00}. These
two approaches have no well-understood region of overlap.  Given the
qualitative differences between the fluid and solid states, a question
that has been open for some time relates to the criterion for
transition between fluid and solid states of granular materials. We
focus on this question.
  
\begin{figure}
  \centering{\includegraphics[width=3.45in]{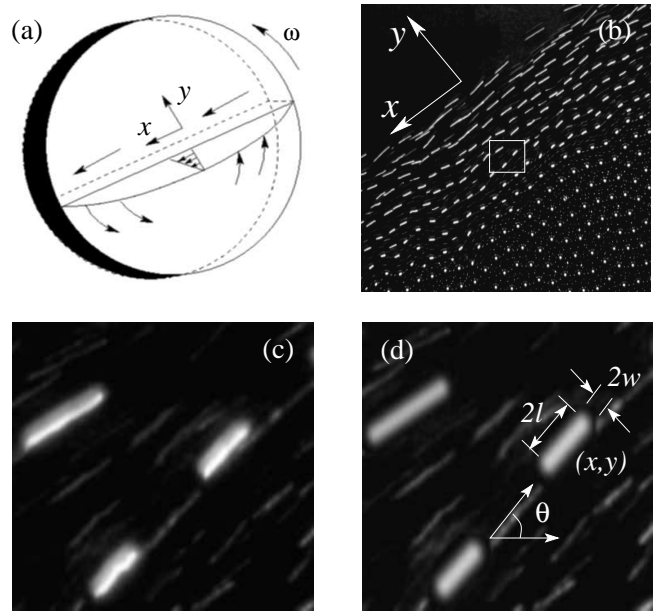}}
  \caption{(a)
    Schematic of the rotating cylinder geometry showing the flowing
    layer and the co-ordinate system employed. (b) A typical image
    showing the portion of the flowing layer at the center of the
    cylinder, with streaks of various lengths formed across the layer
    for a shutter speed of $1/250$ s.  (c) Magnified image of the
    rectangular region marked in (b) showing streaks generated by
    three different particles. (d) The same three streaks with the
    fitted intensity function.  The optimization yields length ($2l$),
    width ($2w$), orientation angle ($\theta$) and position ($x,y$) for
    every streak.}
  \label{fig:schem}
\end{figure}

\begin{figure*}
  \centering{\includegraphics[width=6.5in]{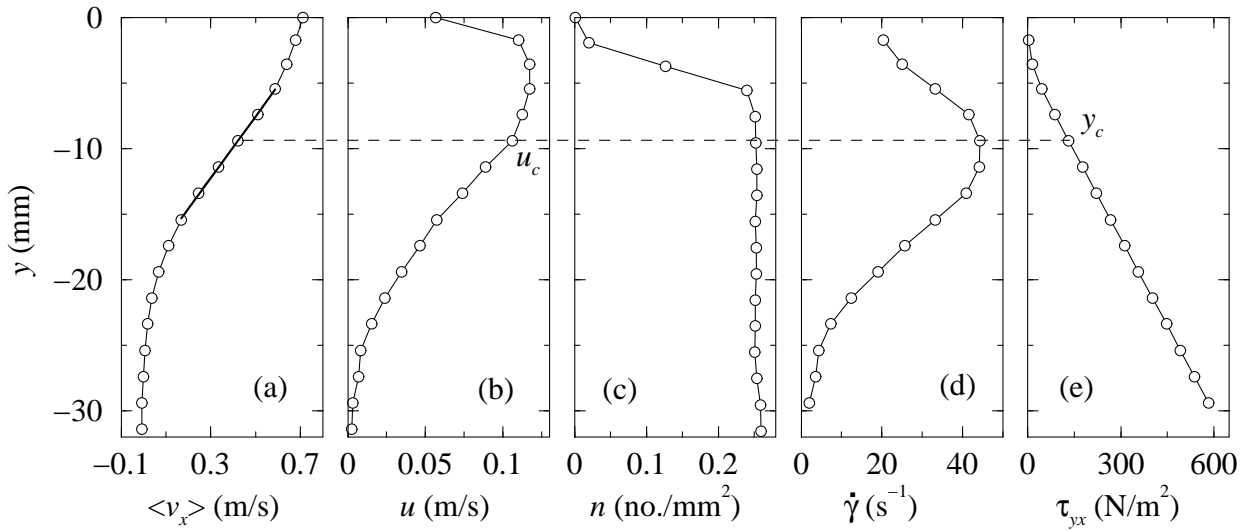}}
  \caption{Mean velocity ($\langle v_{x}\rangle$), rms velocity ($u$), number
    density ($n$), shear rate ($\dot{\gamma}$) and shear stress ($\tau_{yx}$)
    across the flowing layer for $2$ mm steel balls rotated at $3$
    rpm. The solid line in (a) shows a linear fit to the mean velocity
    profile. The dashed line denotes the transition point in all the
    figures.}
  \label{fig:basedata}
\end{figure*}

There are few studies which focus on the transition between the solid
and fluid states. \citet{met02} studied solid to fluid transition in a
horizontally vibrated container of beads. They observed hysteresis in
the transition which was well-predicted by a dry friction model in
which the friction coefficient varies smoothly between a dynamic and
static value. A fluid-solid transition was also observed by
\citet{danna01a} for vertically vibrated particles.  They found that
the vibrating granular medium first undergoes a sharp transition to a
super-cooled liquid and then gradually achieves a solid-like state on
reducing the intensity of vibration. This process follows a modified
Vogel-Fulcher-Tammann (VFT) model which is typical of fragile glasses.

Coexisting solid and fluid phases have been studied primarily in the
context of surface flows; these comprise a layer of fluid like flow on
a fixed bed of the same material. Examples of surface flows which have
been well-studied include heap flows
\citep{bouc94,bout98a,gra99,dou99,kom00,lem00,kha01b} and rotating
cylinder flows \citep{raj90,elp98,orp01,kha01c,jain02,bon02,hill03}.
Remarkably simple theories describe the coexistence between the solid
and fluid. In the simplest versions, the local {\it melting\/} and
{\it freezing\/} is determined by the local angle of the solid-fluid
interface: If the local angle is greater than a ``neutral'' angle, the
solid melts (the heap erodes) so as to reduce the angle and vice versa
\citep{bouc94,bout98a}.  Continuum models based on Coulombic friction
models for the solid region and simple rheological models for fluid
region predict a similar behaviour \citep{dou99,kha01b}.  Experimental
studies show that the models based on local angle based
melting/freezing give good predictions \citep{kha01b}.

An assumption in the coarse-grained models described above is the
existence of an interface between the solid and fluid. However, the
recent work 
of \citet{kom00} indicates that the surface flow on a heap decays
smoothly with depth and generates motion deep within the heap.  The
interface between the solid and fluid regions is thus not well-defined
when considered at a particle length scale.  Studies of the velocity
profile in rotating cylinders flow also confirm this picture of a
smooth decay of the velocity into the bed, rather than an abrupt
change at an interface \citep{jain02,bon02}.

The objective of the present work is to gain an insight into the
fluid-solid transition in granular flows focussing on a system where
both fluid-like and solid-like regions coexist. We find a well-defined
transition point in the system which demarcates two distinct flow
regions. The behaviour in each region is characterized and results
indicate that the transition has some characteristics of a glass
transition.

Experiments are carried out in quasi-2d aluminium cylinders (length
$1$ or $2$ cm) of radius $16$ cm (Fig.~\ref{fig:schem}a). The end
walls are made of glass and a computer controlled stepper motor with a
sufficiently small step is used to rotate the cylinders.
Monodisperse, spherical, shiny stainless steel balls of three
different sizes with diameters ($d$) $1$, $2$, and $3$ mm are used in
the experiments.  Cylinder rotation in the rolling flow regime
(rotational speed, $\omega = 2-9$ rpm) produces a thin flowing surface
layer and measurements are made near the center of the cylinder where
the layer thickness is maximum and the flow is non-accelerating. The
particles are heavy enough and conductive so that charge effects are
negligible.  The experiments are carried out with $50 \%$ of the
cylinder filled with particles.

The motion of the particles is captured by taking high resolution
images using a digital camera (Nikon Coolpix 5000) in the presence of
an incident beam of light. The size of the recorded region is $2560 \times
1920$ pixels, with one pixel corresponding to $0.016-0.03$ mm
depending on the distance of the camera to the cylinder. The point
source of light is directed near parallel to the end wall of the
cylinder so as to illuminate only the front layer of the flowing
particles.  Each moving particle generates a streak of definite length
depending on its speed and the shutter speed of the camera. Images are
taken for a range of camera shutter speeds ($1/15-1/2000$ s) so as to
account for the varying velocity across the flowing layer. This gives
streaks, which are adequately long for analysis but not so long as to
overlap with other streaks, in each the part of the flowing layer
(Fig.~\ref{fig:schem}b). Two hundred images are taken for each shutter
speed with an overall of two thousand images combined over different
shutter speeds. Due to the time interval between photographs, each
experiment typically takes 500 cylinder revolutions.

A parameterized intensity function corresponding to a stretched
Gaussian function is then fitted to the intensity values of the streak
pixels (and an immediate neighbourhood) in the image by means of a
standard non-linear optimization technique. The fitting yields the
length ($2l$), width ($2w$), orientation angle ($\theta$) and the position
($x,y$) for the streak (Fig.~\ref{fig:schem}d). The analysis
technique was calibrated by 
carrying out experiments for a single particle glued to the cylinder
end plate. The error in velocity
measurements was found to be less than than $3\%$.  For analysis, the
flowing layer region is divided into bins of width equal to the
particle diameter and length $20$ mm parallel to the flowing layer.
The components of the mean and the root mean square (rms) velocities
for each bin are calculated by averaging over all streaks in a bin.

Fig.~\ref{fig:basedata} gives the variation of system variables with
depth ($y$) in the flowing layer for $2$ mm particles. The mean
velocity ($\langle v_{x}\rangle$) profile is smooth and shows 3 regions; a
near-linear middle region, an exponentially decaying region at the
bottom and a flattened region near the top. Similar profiles have been
reported in several previous studies
\citep{boc01,longo02,bon02,jain02,mueth03,hill03,tab04,mueth00}. The
flattened upper region which is the very low density region
corresponding to saltating particles is not seen in some studies. The
magnitudes of the rms velocities in the two directions are different:
the component in the flow direction ($x$-component) is about 10\%
higher than that in direction perpendicular to the flow
($y$-component).  The profile of the total magnitude of the
fluctuating velocities ($u$) is shown in Fig.~\ref{fig:basedata}b.
The rms velocity profile shows two distinct regions: a relatively slow
variation near the free surface followed by a sharper decrease deeper
in the bed. Fig.~\ref{fig:basedata}d shows the shear rate variation in
the layer, obtained by numerical differentiation of the data in
Fig.~\ref{fig:basedata}a.  The shear rate increases to a maximum value
(corresponding to the inflection point in the velocity field) and then
decreases. We note that an oscillating shear rate profile is obtained
if a smaller bin size is used as reported previously
\citep{mueth00,mueth03,hill03}. 
The transition point in the fluctuation velocity profile
coincides with the maximum in the shear rate. The areal number density
is almost constant throughout the flowing layer with a rapid decrease
near the free surface as shown in Fig.~\ref{fig:basedata}c. 

\begin{figure}
  \centering{\includegraphics[width=3.0in]{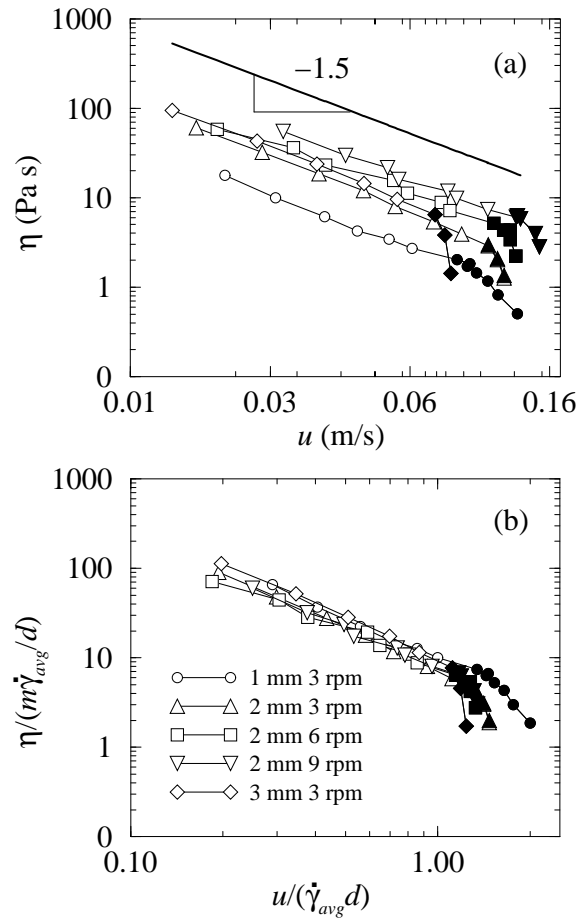}}
  \caption{Apparent viscosity ($\eta$) variation with rms velocity ($u$)
    for all the cases studied. Filled symbols represent the points at
    and above the transition velocity ($u_{c}$). (a) The solid line at
    the top represents a linear fit (slope $\simeq{1.5}$) to the data below
    the transition point for all the cases studied. (b) The data is
    scaled using particle mass ($m$), particle diameter ($d$) and
    average shear rate obtained by a linear fit to each corresponding
    velocity profile.}
  \label{fig:eta_u}
\end{figure}

\begin{figure}
  \centering{\includegraphics[width=3.0in]{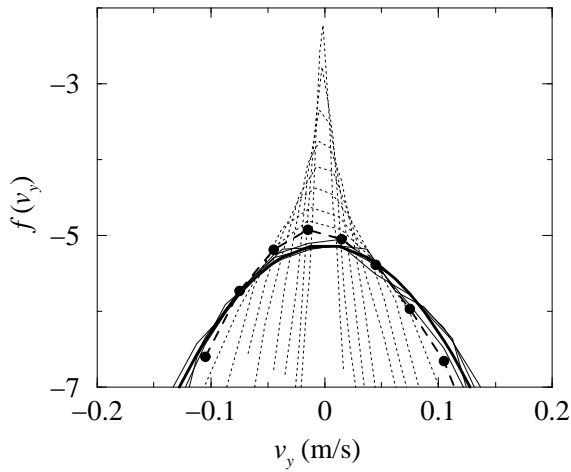}}
  \caption{Distributions of velocity in $y$ direction at
    various locations across the layer for $2$ mm steel balls rotated
    at $3$ rpm. The dashed line with filled circles represents the
    transition point, the solid lines represent the region above the
    transition point and all the dotted lines represent the region
    below the transition point. The thick black line represents a
    fitted parabola to the curves at and above the transition point.}
  \label{fig:vydist_layer}
\end{figure}
For
non-accelerating flows a force balance yields the shear stress as
$d\tau_{xy}/dy = \rho g \sin\beta \approx \rho_{b} (n/n_{b}) g \sin\beta$ where $g$ is
acceleration due to gravity, $\beta$ is the angle of repose, $n$ is the
number density in the flowing layer, $n_{b}$ is
the number density in the rotating packed bed and $\rho_{b}$ is the bulk
density of the rotating packed bed. We neglect the contribution of wall
friction in the estimate. This could become significant deeper in the
bed \citep{tab04}. However, based on the method of \citet{tab04} we
find that the contribution is about $10\%$ of the total stress in the
cases studied and does not qualitatively affect the results. Upon
integration we obtain $\tau_{xy} = (\rho_{b}/n_{b}) g \sin\beta \int_{y}^{0} n dy$
assuming $\tau_{xy} = 0$ at the free surface. The shear stress shows a
near linear increase with depth (Fig.~\ref{fig:basedata}d). The
results of Fig.~\ref{fig:basedata} indicate that there is a
qualitative change in the rheology at the transition point ($y_{c}$).
Above $y_{c}$, the shear stress increases with shear rate which is
typical of fluids. However, below $y_{c}$ the shear stress increases
while the shear rate decreases. This implies that the viscosity
increases sharply with depth below $y_{c}$, even though the number
density is nearly constant.

Fig.~\ref{fig:eta_u}a shows the variation of the apparent viscosity
($\eta = \tau_{xy}/\dot{\gamma}$) with the rms velocity ($u$), considering only
the region of constant density. The data points at and above the
transition point ($y\geq y_{c}$) are plotted as filled symbols. There is
a sharp transition in this case as well and the transition occurs at
the same value of $u$ as in Fig.~\ref{fig:basedata}b.  In the region
near the free surface there is a rapid increase in viscosity with
decreasing rms velocity whereas in the region approaching the fixed
bed there is a much slower power law increase with an exponent
$n\sim-1.5$ for all the cases studied. The data scales with
$\dot{\gamma}_{avg}$ and $d$ to fall on a single curve
(Fig.~\ref{fig:eta_u}b), and the transition
occurs at $u_c\sim 1.2\dot{\gamma}_{avg}d$ for all cases. Here $\dot{\gamma}_{avg}$
is the shear rate obtained by fitting a straight line to the linear
portion of the velocity profile (Fig.~\ref{fig:basedata}a).

Fig.~\ref{fig:vydist_layer} shows the distributions of the $y$
direction velocities at different locations in the layer. The $v_y$
distribution is Gaussian for all points above the transition point
($y\geq y_{c}$) and it gradually evolves
to a Poisson distribution as we go deeper into the bed. A Poisson
velocity distribution was also found by \citet{mueth03} for Couette
flow of a dense granular material. The $v_x$ distribution (not shown)
is Gaussian above the transition point. However, below the transition
point the behaviour is complex and bimodal distributions are obtained.

The results presented show a sharp transition between two flow
regimes. In the upper region near the free surface, the behaviour is
fluid-like and the velocity distributions are Maxwellian. Below the
transition point the material appears to be an amorphous soft solid,
increasing in strength with depth in the layer. The transition to this
solid-like regime occurs at a relatively large rms velocity ($\approx 0.1$
m/s). We conjecture that the sharp transition occurs because of the
formation of a percolated network of particles in extended contact
with each other. This is in contrast to the fluid-like regime where
the particles interact through collisions. The contact network coexists
with fluid-like domains and the fraction of particles which are part
of the network increase with depth. The transition appears to be
analogous to a glass transition. New rheological models may be
required for granular flows that span these two regimes.

\bibliography{/home/ash/rpcs/thesis/ref}

\end{document}